\documentstyle[12pt,aasms4]{article}
\begin{document}

\title{Infrared Variation of Radio Selected BL Lacertae Objects}

\author{J.H. Fan}

\affil{ CCAST (World Laboratory), P.O. Box 8730, Beijing 100080, China 
 \and Center for Astrophysics, Guangzhou Normal University, Guangzhou 510400, 
 China, e-mail: jhfan@guangztc.edu.cn}
\and
\author{R.G. Lin}
\affil{ Center for Astrophysics, Guangzhou Normal University, Guangzhou 
 510400, China}
\date{Received <date>;accepted<date>}

\begin{abstract}

 In this paper, the historical infrared (JHK) data compiled from the published
 literature are presented in electronic form for  40 radio selected BL 
 Lacertae objects (RBLs)  for the first time. Largest variations are found 
 and compared with the 
 largest optical variation. Relations between color index and magnitude 
 and between color-color indices are discussed individually. For the 
 color-magnitude relation, some objects ( 0048-097,  0735+178, 0851+202, 
 1215+303, 1219+285,  1749+096, and perhaps 0219+428,  0537-441, 1514-241) 
 show that color index increases with magnitude indicating that the 
 spectrum flattens when the source brightens while some other objects
 ( 0754+100, 1147+245, 1418+546, and 1727+502) perhaps show an opposite 
 behaviour, while remaining objects do not show any clear tendency; 
 For color-color relation,  it is found  common that (J-K) is closely 
 correlated with (J-H) while (J-H) is not  correlated with (H-K). 
 \end{abstract}

 \section{Introduction}
 While the nature of active galactic nuclei (AGNs) is still an open problem, 
 the study of AGNs variability  can yield valuable information about 
 their nature.  Photometric observations of AGNs are important to 
 construct light curves and to study  variability behavior over
 different time scales. In AGNs, some long-term optical variations have been 
 observed and in some cases claimed to be periodic (see Fan et al. 1998a).

 Before $\gamma$-ray observations were available, Impey \& Neugebauer (1988) 
 found that the infrared emission  (1-100 $\mu m$) dominates the bolometric 
 luminosities of blazars. The infrared emissions are also an  important 
 component for the luminosity even when the $\gamma$-ray emissions are 
 included ( von Montigny 1995). Study of the infrared will provide much 
 information of the emission mechanism. The long-term optical variations 
 have been discussed in the papers: Webb et al. (1988); Pica et al. (1988); 
 Bozyan et al. (1990); Stickel et al. (1993); Takalo (1994); Heidt \& Wagner 
 (1996) and references therein. The long-term infrared variations have been 
 presented for some selected blazars in a paper by Litchfield et al. (1994).

 BL Lacertae objects are an extreme subclass of AGNs showing rapid and large 
 amplitude variability, high and variable polarization, no or weak emission
  line features (EW $ < 5\AA)$. From surveys, BL Lacertae objects can be 
 divided into radio selected BL Lac objects (RBLs) and X-ray selected BL 
 Lac objects (XBLs); From the differences in their overall spectral energy 
 distributions as represented by their location in the  $\alpha_{ro}$ versus 
 $\alpha_{ox}$, they can be divided into quasar-like BL Lacertae objects 
 (Q-BLs) and X-ray strong BL Lacertae objects (X-BLs) (Giommi et al. 1990); 
 From the difference in the peak frequency of the spectral energy 
 distribution (SED), they can be divided into high frequency peaked BL Lac 
 objects (HBLs) and low frequency peaked BL Lac objects (LBLs) (Padovani 
 \& Giommi 1995; Urry \& Padovani 1995).  In the last catalogue of active 
 galaxies, Veron \& Veron (1998) list 357 BL Lac objects and BL Lac object 
 candidates. 

 There are some obvious differences between  RBLs and  XBLs (see Fan 1998 
 in preparation, for a brief review). Some authors claim that they are the 
 same type of objects with XBLs 
 having wider viewing angle than RBLs (Ghisellini \& Maraschi, 1989; 
 Georganoppoulos \& Marscher 1996; Fan et al. 1997); Other claim 
 that there is a continuous spectral sequence from XBLs to RBLs (Sambruna 
 et al. 1996) while some claim that they are different classes 
 (Lin et al. 1997).

 Infrared observations have been done for BL Lac objects for more than 20 
 years. Observations in the infrared were obtained originally on BL Lacertae 
 by Oke et al. (1969). Strittmatter et al. (1972) observed the continuity 
 of spectral-flux distribution from the optical to infrared wavelengths. 
 Afterwards, some other authors also did similar observations (see Stein 
 1978 for summary).  Using the 1.26-m infrared telescope at Xing Long 
 station, Beijing observatory, we obtained some infrared observations of a 
 few BL Lac objects ( Xie et al. 1994; Fan et al. 1998b). But long-term 
 infrared variations are not available in the literature. 
 In this paper, we mainly present the long-term infrared (J,H, and K bands ) 
 light curves for some RBLs and discuss their variability properties.
  The paper has been arranged as follows: In section 2, we review the 
 literature data and  light curves; in section 3, we discuss the 
 variations and give some remarks for each object,  and a short  summary.

\section{Near-infrared Light Curves}

\subsection{Data from Literature }

 Infrared observations are available since the end of the 1960s. We 
 compiled the data  from 58 publications. They are listed in table 1 \& 2,
  which gives the observers in Col. 1.  and the telescope(s) used in Col.2. 
 For the J, H, and K magnitudes, they are listed in an table 4 in
 electronic form. In table 4, Col. 1 gives the name; Col. 2. the JD time;
 Col. 3, the J magnitude; Col. 4 the uncertainty for J; Col. 5, the 
 H magnitude; Col. 6, the uncertainty for H; Col. 7, K magnitude;
 Col. 8, the uncertainty for K.

\subsection{Data Analysis}

 The flux densities from the literature had been converted back to magnitudes
 using the original formulae.  In the literature, different telescopes 
 are used, different telescopes use photometers with slightly
 different filter profiles, resulting in slightly  different calibration 
 standards and zero-points, but the  uncertainty aroused by the different 
 systems is no more than a few percent.  

 The magnitude is dereddened using  
 $A(\lambda)=A_{V}(0.11\lambda^{-1}+0.65\lambda^{-3}-0.35\lambda^{-4})$
 for $\lambda > 1 \mu m$ and  $A_V$ = 0.165(1.192- $\tan{b})\csc{b}$ for 
 $ |b| \leq 50^{\circ}$ and $A_V$ = 0.0 for  $ |b| > 50^{\circ}$ 
 (Cruz-Gonzalez \& Huchra 1984, hereafter CH (84); Sandage 1972; 
 Fan et al. 1998c).  It is clear that some RBLs have many observations 
 while others have only a few data points, in Figures 2-22, we  show only 
 those with  more observations.

 For data shown in Figs. d - i of each object we have
 performed linear fit with uncertainties  in both coordinates 
 considered (see Press et al. 1992 for detail), 
 \begin{equation}
 y(x) = a + bx
 \end{equation}
 In principle, $a$ and $b$ can be determined  by 
 minimizing the $\chi^{2}$ merit function (i.e. equation 15.3.2 in 
 Press et al. book)
 \begin{equation}
\chi^{2}(a,b)= {\sum\limits_{i=1}^{n} {(y_{i}-a-bx_{i})^{2}w_{i} }} 
 \end{equation}
 where ${\frac{1}{w_{i}}}=\sigma^{2}_{y_{i}}+b^{2}\sigma^{2}_{x_{i}}$,  $\sigma_{y_{i}}$ and $\sigma_{x_{i}}$ are the $x$ and $y$ standard 
 deviations for the $i$th point
 Unfortunately, the occurrence of $b$ in the denominator of the above 
 $\chi^{2}$ merit function, resulting in equation
 ${\frac{\partial \chi^{2}}{\partial b}}=0$ being
 nonlinear makes the  fit very complex, although we can get a 
 formula for $a$ from
  ${\frac{\partial \chi^{2}}{\partial a}}=0$:
 \begin{equation} 
  a={\frac{\sum\limits_{i}w_{i}(y_{i}-bx_{i})}{\sum\limits_{i}w_{i}}}
 \end{equation}
  Minimizing the $\chi^{2}$ merit function, equation (2), with respect to 
 $b$ and using the equation (3) for $a$ at each stage to ensure that the 
 minimum with respect to $b$ is also minimized with respect to $a$, we 
 can obtain $a$ and $b$. As Press et al. stated, the linear correlation 
 coefficient $r$ then can be  obtained by  weighting the relation (14.5.1)
  terms of Press et al. (1992).  
 Finding the  uncertainties, $\sigma_{a}$ and $\sigma_{b}$, in $a$ and $b$ is
 more complicated (see Press et al. 1992). Here, we have  not performed this.
 For the data whose uncertainties were not given in the original literature 
 are not included in our linear fit or in Fig. d-i, but they 
 are included in the light curves (Fig. a - c).

 As discussed in the paper of Massaro \& Trevese (1996), there is a 
 statistical bias in the  spectral index-flux density correlation. Following 
 their suggestion, we considered the relation between magnitude in one 
 band and the color index obtained fro two other  bands to avoid this
 bias. The relations are shown in Fig. d-f.

\section{Discussion}

\subsection{Variations}

 BL Lac objects are variable at all wavelengths. For the RBLs with more 
 infrared data, we presented their variability properties in Table 3, which 
 gives the name in Col. 1, redshift in Col. 2, Reddening correction, $A_{V}$,
 in Col. 3, largest optical polarization ($P_{opt}$) in Col. 4, largest 
 amplitude optical variation ($m_{opt}$) in Col. 5, largest amplitude 
 variations in J, H, and K in Col. 6, 7, and 8, the averaged color index 
 (J-H) and (H-K) in Col. 9, and 10, the uncertainty in Col. 9  and 10 
 is the one sigma 
 deviation. The infrared variations are very large for some objects. 
 The color indices are also variable and associated with the magnitude 
 for some objects,  indicating that the spectrum flattens when the source 
 brightens while some objects show different behaviours. For 
 variation, it is reasonable for larger variation to correspond to shorter 
 wavelength, but this paper shows that some objects do show a different 
 behaviour. 
 The objects showing large optical variation also show large infrared 
 variations and high polarizations.  Large variation and high polarization 
 are associated. For the color-indices, it is common that there is a 
 correlation for (J-K) and (J-H) while there is almost no correlation for 
 (J-H) and (H-K) although some objects show different properties.
 But, when the averaged color indices are taken into account, there are 
 mutual correlations between any two color indices of (J-H), (J-K) and 
 (H-K) with the correlation between (J-H) and (H-K) being not so
 close as compared with other two correlations. (J-K) = 2.26(J-H) - 0.28 
 with correlation coefficient  $r$ = 0.90 and the probability of chance
 correlation  $p = 6.3\times 10^{-11}$, (H-K) = 1.44(J-H) - 
 0.41 with $r$ = 0.80 and $p = 4.4\times 10^{-7}$, and (J-K) = 1.62(H-K) + 
 0.33  with $r$ = 0.93 and $p = 6.1\times 10^{-13}$. The (J-H) and (H-K) 
 relation is shown in Fig. 1.  In the next section, individual objects 
 are discussed.

\subsection{Remarks}
\subsubsection{PKS 0048-097, $OB-081, MC3, PHL856$}

 There is no redshift available for this polarized ( $P_{3.7cm}=9.3\%$, 
 Wardle 1978;  $P_{opt}=$1$\%$-27 $\%$, CH(84);  Brindle et al. 1986 ) 
 object. Very faint infrared  fluxes have been observed by Gear (1993), 
 which leads 
 the object to have the largest infrared variation among the objects 
 discussed in the present paper. The infrared variations  are $\Delta J$ = 
 6.55,  $\Delta H$ = 6.23, and $\Delta K$ = 6.07 while that  in the 
 optical band is 2.7 mag (Angel \& Stockman 1980). We can expect that the 
 optical  monitoring should get a similar amplitude.  For 
 color-color indices,  there are correlations for (J-K) and (J-H) and for 
 (J-K) and (H-K):  
 (J-K) = 1.20(J-H) + 0.61 with $r$ =0.69 and  $p = 5.\times 10^{-4} $, 
 (J-K) = 4.7(H-K) - 2.13  with $r = 0.64$ and $p = 2.\times 10^{-3}$, 
 but no  correlation was found for (J-H) and (H-K). For the color indices 
 and the  magnitudes,  there are correlation indications of (H-K) increasing 
 with J ($p \sim 1\%$) and  (J-K) with H ($p \sim 0.3\%$) (see Fig 2d, 2e).

\subsubsection{PKS 0118-272}

 An absorption feature at $z$ = 0.559 was found by Falomo (1991). It was 
 observed  in the infrared during the period JD 244000+ 4773-7753. For 
 this object,  there are no correlations  for color-color indices or for 
 color index and  magnitude.  It is a highly polarized object 
 ( $P_{opt} \sim 17 \%$) and  shows $ dP/d\nu > $ 0 (Mead et al. 1990).

\subsubsection{PKS 0138-097}

 It was observed by Falomo et al (1993) at the visual and 2.2$\mu m$ 
 wavelengths,  the spectral flux distribution is well described by a single
  power law of 
 $\alpha$ = 1.0 ($f_{\nu} \sim \nu^{-\alpha}$). It shows variable and high
 optical polarization ($P = 6\%-29.25\%$)(Mead et al. 1990). The variation in 
 the infrared is almost the same as that in the  optical band ($\Delta m$ = 
 1.52 on a time scale of one week, Bozyan et al. 1990). 
  There is a correlation for (J-K) and (J-H) but not for other colors.

\subsubsection{0219+428, $3C 66A$}

 A redshift of 0.444 has been derived from a possible  MgII  $\lambda$2800 
 feature (Miller et al. 1978). A rapid decline of 1.2 mag in 6 days was 
 observed by Takalo et al. (1992). Different properties for the color and 
 brightness are found in this source: Takalo et al. (1992) found that 
 although the spectral index is considerably flat when the source was at 
 its observed maximum state, the (J-H) and (H-K) against K color-magnitude 
 diagrams are essentially scattered; Massaro et al. (1995) found that there
 is a positive correlation between the infrared spectral index and the 
  brightness at more than the 98 $\%$ level while De Diego et al. (1997) 
 found a different behaviour.  From the long-term data compiled in the 
 present paper no clear correlation has been found for color index
 and magnitude except that there is a tendency for (H-K) to increase with 
 J (Fig. 3e). There are close correlations for (J-K) and (J-H) (Fig. 3h) 
 and for (J-K) and (H-K) (Fig. 3i) but not for (J-H) and (H-K):
 (J-K) = 1.38(J-H) + 0.33 with $r = 0.76$ and $p = 10^{-7}$, 
 (J-K) = 1.60(H-K) + 0.18 with $r = 0.45$ and $p = 5.0 \times 10^{-3}$.
 It shows two lower states in the light curves (Fig. 3 a,b,c). Its 
 largest variation of 2.0 mag (Miller \& McGimsey 1978) obtained in the 
 optical band is similar to those found in the infrared band (1.67mag in 
 the K  for instance). Polarizations of $P_{opt}$ = 6$\%$-18$\%$ and 
 $P_{Rad.}$ = 5$\%$ are reported in the papers of (CH(84)) and Takalo (1991).

\subsubsection{AO 0235+164}

 Absorption redshifts of $z_{abs}$ = 0.524 and 0.852 were found by Burbidge 
 et al. (1976) and an emission line redshift of $z_{em} = 0.94$ was found by 
 Nilsson et al.  (1996). It is one of the most extensively studied objects 
 in the infrared  (Impey et al. 1982; Gear et al. 1986a; Takalo et al. 1992), 
 variations of  $\Delta J$ = 1.9, $\Delta H$ = 1.8, and $\Delta K$ = 1.9 are
  found by Takalo et al. (1992).  The spectral index is found to be 
 correlated with J flux at a significance level of 
 more than 90$\%$ by Brown et al. (1989). It is the most highly  polarized 
 ($P(V) = 44 \%$,$P_{IR} = 36.4 \%$) and the reddest ($\alpha$ = 4.61) object 
 (Impey et al. 1982; Stickel et al. 1993).  The infrared light curves show 
 three clear outbursts, and variations  as large as the optical ones 
 ( $\Delta m$ = 5.3 mag, Stickel et al. 1993). There are positive correlations 
 between color indices against magnitude when it is brighter than J = 14., 
 H = 12.5 and K = 12.0 magnitudes, but the correlation is complex in fainter 
 states. There are correlations for (J-K) and (J-H) and 
 (H-K): (J-K) = 1.30(J-H) + 0.64 with $r$ = 0.56 and $p = 3 \times 10^{-5}$,
 (J-K) = 1.42(H-K) + 0.56  with $r$ = 0.53 and $p = 9 \times 10^{-5}$. 
 (see Fig. 4)

\subsubsection{0300+471, $4C47.08$}

 This object has been observed in the infrared  only once  (Puschell 
 \& Stein 1980) with H = 13.44 and K = 12.46, which gives a color index of 
 (H-K) = 0.98$\pm$ 0.07.  Its flux distribution from 0.36$\mu m$ to 2.3$\mu m$ 
 can be expressed by a power law with  $\alpha=1.0$ 
 ( $f_{\nu} \propto \nu^{-\alpha}$). It is  11.5$\%$ polarized at 3.7cm 
 (Wardle 1978) and 24$\pm$5$\%$ polarized at 0.44$\mu m$ (Puschell \& Stein 
 1980).

\subsubsection{0306+103}

 This variable ($\Delta m=2.2$ mag, Leacock et al 1976) object has been  
 observed in K for only two times (K =11.82 and 11.92 mag).

\subsubsection{PKS 0537-441}

 Large variation of $\Delta m$ = 5.4 mag (Stickel et al. 1993) in the 
 optical and a variation of $\Delta H$ = 2.0 mag (CH(84)) are found in 
 this highly polarized ($P_{opt} = 18.7\%$, Impey \& Tapia 1988) object. 
 From our compiled data, a large variation of 3 mag in J and weak 
 correlations  between  (J-K) and  (J-H) and (H-K) are found:
 (J-K) = 1.92(J-H) + 0.04 with $r$ = 0.54 and $p = 2.7\%$,
 (J-K) = 1.42(H-K) + 0.47 with $r$ = 0.547 and $p = 2.7\%$. 
 Figs. 5d,e show that there is a tendency for the color index to increase
 with the magnitude.

\subsubsection{PKS 0735+178, $VRO 17.07.02, OI158$}

 It has been observed in the optical band for about 90 years (Fan et al. 
 1997) and in the infrared for about 20 years (Lin \& Fan 1998). The largest 
 optical variation of $\Delta m = 4.6$ (Fan et al. 1997) is greater than 
 that in the infrared (Lin \& Fan 1998). Rapid variability of 0.4 mag in 
 about 15 minutes and a  positive correlation between infrared spectral 
 index and brightness are found (Massaro et al. 1995; Brown et al. 1989). 
 It shows correlations between H and (J-K) and between K and (J-H) for 
 long-term data: (J-K) = 0.12H + 0.17 with $r$ = 0.447 and 
 $p \sim 5.3 \times 10^{-5}$);
 (J-H) = 0.04K + 0.32 $r$ = 0.30 and $p \sim 7.8 \times 10^{-3}$ ) 
 (Fig. 6d and 6e, see also Lin \& Fan 1998). (J-K) is found to be correlated 
 with both (J-H) and (H-K) (see Lin \& Fan 1998).
 It is highly polarized in the radio ($P_{3.7cm} = 4.7\%$, Wardle 1978), 
 in the infrared ($P_{IR} = 32.6\%$, Impey et al. 1984) and in the optical 
 ($P_{opt}=35\%$, see Lin \& Fan 1998) bands.

\subsubsection{PKS 0754+100, OI 090.4}
 
 PKS 0754+100 is  observed to show rapid variability in the infrared 
 ($P_{IR} = 4\%-19\%$)  and in the visual  ($P_{Opt.} = 4-26 \% $) 
 polarizations, a variation of  $6\%$ over 4 hours in optical polarization 
 was observed by Impey et al. (1982, 1984).  The polarization is independent 
 of wavelength between the visual and the  2.2$\mu m$ regions 
 (Craine et al. 1978). An optical variation  of 2.0 magnitude 
 was found from the examination of the plate archives by Baumert (1980) 
 and a variation of 1 mag in the infrared was reported (CH84). It is the only 
 object in Massaro et al (1995) sample which, despite a recorded 
 variability of  about 1 mag, does not show statistically significant 
 changes of the spectral index.  
 0754+100 shows a decreasing brightness since the first infrared observation
 although there are some brightness fluctuations in the light curves.  It is 
 interesting that the variation in H is greater than those in J and K with 
  variability ranges of J = 12.31 to 14.08, H = 11.21 to 13.27 and 
 K = 10.55 to 12.43. Besides, although an anti-correlation between H and (J-K) 
 appears in Fig. 7d, it is far from any firm conclusion. (J-K) is 
 correlated  with both (J-H) and (H-K):
 (J-K) = 1.3(J-H) + 0.54 with $r$ = 0.78 and $p = 2.4 \times 10^{-8}$,
 (J-K) = 2.16(H-K) - 0.15 with $r$ = 0.61 and $p = 8.9 \times 10^{-5}$.

\subsubsection{S4 0814+42, $OJ425$}

 This object is observed to show variations of 1.62, 1.63, and 1.52 mags 
 in J, H, and K respectively (Gear 1993). Mean color indices of (J-H)
  = 0.96$\pm$0.05, (J-K) = 1.90$\pm$0.04, and (H-K) = 0.95$\pm$0.07 can 
 be obtained from those observations. Polarizations of $P_{opt} = 12 \%$ 
 (Impey et al. 1991) and $P_{3.7cm}=9.8$ (Wardle 1978) are reported.

\subsubsection{PKS 0818-128, OJ-131}

 It shows optical variation from 14.6 to 17.5 magnitude. The 
 linear polarization is high and variable with $P_{opt} = 8\%$-36$\%$ 
 (Craine et al. 1978) and $P_{IR} = 9\%-14\%$ (Impey et al. 1984). 
 The infrared variation of 2.0 mag is a little smaller than  in the 
 optical band ($\Delta m = 2.9$, Craine et al. 1978). There are indications 
 of correlations between (J-K) and (J-H) and (H-K), but no definite 
 conclusion can be drawn from the few points with large error bars. The 
 correlation between the color index and magnitude is also complex as
 shown in Fig 8.

\subsubsection{PKS 0820+22, $4C22.21, MG0823+2222$}

 This RBL was observed twice and does not show any clear variation in the 
 infrared: J = 17.63$\pm$0.16-17.76$\pm$0.07, H = 16.54$\pm$0.07, and 
 K = 15.59$\pm$0.04-15.75$\pm$0.14 (Gear 1993). Averaged color indices of 
 (J-H) = 1.22$\pm$0.10, (J-K) = 2.02$\pm$0.17, and (H-K) = 0.95$\pm$0.08 
 can be obtained. It is at a redshift of 0.951 (Stickel et al. 1991) and 
 $P_{opt} = 5.2\%$ polarized in optical band (Kuhr \& Schmidt 1990).

\subsubsection{PKS 0823+03, $OJ038, MG0825+0309$}

 This highly polarized ($P_{opt}=22.9\%$, Impey \& Tapia 1988) object shows 
 a variation of about 0.8 mag in the infrared: J = 13.73-14.49, 
 H = 12.98-13.79, and K = 12.39-13.24 (Gear 1993).  Mean color indices are
 (J-H) = 0.72$\pm$0.04, (J-K) = 1.29$\pm$0.07 and (H-K) = 0.57$\pm$0.03.  
 Its redshift is 0.506 (Stickel et al. 1991).

\subsubsection{S4 0828+493,$OJ448, BP077$}

 An infrared variation of about 0.5 mag (J = 15.81-16.39, H = 14.80-15.44, 
 K = 14.00-14.53) is observed (Gear 1993) from this polarized ($P_{opt} 
 = 7.9\%$,  Kuhr \& Schmidt 1990) object.  Mean color indices are
  (J-H) = 0.98$\pm$0.04,  (J-K) = 1.79$\pm$0.04, and (H-K) = 0.81$\pm$0.14. 
 The recorded  optical variation is $\Delta R$ = 2.0 $mag$ 
 (Bozyan et al. 1990).

\subsubsection{PKS 0829+046, $OJ 049$}

 It is also an extremely variable object ( $\Delta m$ = 5.0 Angel \& 
 Stockman 1980).  Polarizations of $P_{3.7cm}=4.8\%$ ( Wardle 1978) and 
 $P_{opt} = 12\%$ (CH(84))  are reported. It is hard to draw any conclusion 
 for the relation between color  index and  magnitude.  It shows a
  correlation between (J-K) and (J-H):
 (J-K) = 1.32(J-H) + 0.48 with $r$ = 0.76 and $p = 4.3 \times 10^{-5}$.
 Between  (J-K) and (H-K), a close correlation with $r$ = 0.68 and 
 $p = 5.4\times 10^{-4}$ can be obtained once the outlying point is excluded
 (see Fig. 9).

\subsubsection{PKS 0851+202, $VRO 20.08.01, OJ287$}

 OJ287 is one of the very few AGNs for which continuous light curve over more 
 than one hundred years has been observed.  A redshift of $z$ = 0.306 is 
 derived from  [OIII]$\lambda$5007 by Miller et al. (1978) 
 and confirmed by Sitko \& Junkkarinen (1985). Its observed properties from 
 the radio to the X-ray have been reviewed by Takalo (1994). OJ287 shows 
 a 12-year  period in the optical (Sillanpaa et al. 1988; 
 Kidger et al. 1992)  and in the infrared (Fan et al. 1998b). It is highly 
 polarized in the radio  ($P_{3.7cm} = 16.6\%$, Wardle 1978), infrared 
 ($P_{IR} = 14.3\%$, Impey et al.  1982) and optical ($P_{opt} = 37.2\%$, 
 Smith et al. 1987) bands.  Its optical variation of 6.0 $mag$ (Fan et al. 
 1998b) is larger than the  infrared variation. There is a tendency for the 
 color index to increase  with magnitude. (J-K) is correlated with 
 both (J-H) and (H-K):
 (J-K) = 1.62(J-H) + 0.35  with $r$ = 0.76, 
 (J-K) = 1.54(H-K) + 0.35  with $r_{s}$ = 0.65 (see Fig. 10 and Fan et al. 
 1998b).

\subsubsection{B2 1101+384, $Mkn 421,  OM303$ }

 This object is found to have nebulosity with a redshift of 0.0308 by 
 Ulrich et al. (1975). A large optical variation of 4.6 mag was observed 
 (Stein  et al. 1976). The observed optical, infrared 
 and radio polarizations are, respectively,  0$\%$-7.9$\%$ (Rieke et al. 
 1977), 2$\%$-4$\%$ (CH(84)) and $3.7\%$ (Wardle 1978). 
 (J-K) is correlated with (J-H): 
 (J-K) = 0.72(J-H) + 0.69 with $r$ = 0.68 and $p = 6.1 \times 10^{-4}$. 
 (H-K) shows an anti-correlation with (J-H), (H-K) = -0.68(J-H) + 0.94 
 ($p = 6.0 \times 10^{-3}$).  For (J-K) and (H-K), it is hard to draw a 
 definite conclusion as  shown in Fig.11i.  No definite color index-magnitude 
 relation can be obtained. The infrared variation  found here is much 
 smaller than the optical band one.

\subsubsection{PKS 1144-379}

 It is a high-redshift object with $z$ = 1.048 ( Stickel et al. 1989). An 
 optical  variation of 1.92 mag  (Bozyan et al. 1990) and an optical 
 polarization of $P_{opt} = 8.5\%$  (Impey \& Tapia 1988) are reported. 
 The infrared variation is larger at lower frequency 
 and is greater than in the optical.

\subsubsection{B2 1147+245, $OM280, MG1150+2417$}

 No redshift is published  for this source. Similar infrared ($P_{IR} = 
 14.8\%$)  and optical ($P_{opt} = 1.5\%-13\%$) polarizations (Impey et al. 
 1982), and  an  optical variation of 0.77mag (Pica et al. 1988) are reported. 
 The variation in H is found greater than those in J and K. Fig. 12f shows  
 that the  spectrum flattens when the source dims, a similar result was 
 also found by Lorenzetti et al. (1989). (J-K) is associated 
 with (J-H):
 (J-K) = 1.2 (J-H) + 0.67 with $r$ = 0.90 and $p = 1 \times 10^{-6}$
 (see Fig. 12).

\subsubsection{B2 1215+303, $ ON325$}

 For this polarized ($P_{3.7cm} = 6.3\%$ and $P_{opt} = 14\%$, Wardle 1978) 
 object, we found (J-H)  to be correlated with  K:
 (J-H) = 0.24K - 2.13  with $r$ = 0.59 and $p = 7.2 \times 10^{-3}$. The 
 infrared variation found here is smaller than that in the optical band 
 ($\Delta m$ = 3.3 mag, Zekl et al. 1981). (J-K) is correlated with (J-H) and 
 (H-K): (J-K) = 1.06(J-H) + 0.60 with $r$ = 0.69 and $p = 1.2\times 10^{-3}$, 
 (J-K) = 1.68(H-K) + 0.27 with $r$ = 0.42  and $p = 5.3\%$. (see Fig. 13)

\subsubsection{B2 1219+285, $W Com, ON231, VR 28.12.02, GC$}

 This extremely variable object ($\Delta V$ = 5 mag) was observed to be 
 polarized at $P_{3.7cm} = 5.8\%$ (Wardle 1978) and  $P_{opt} = 2\%-10\%$ 
 ( CH(84)). A correlation between H and (J-K) was found: (J-K) = 0.28H - 2.01 
 with $r$ = 0.62 and $p \sim 1\%$. A correlation is also found between (J-K)
  and (J-H): (J-K) = 1.04(J-H) + 0.72 with $r$ = 0.61 and $p = 1.1\%$ 
 (see Fig. 14h). 
 For (J-K) and (H-K), there is a close relation if the two outlying points 
 are excluded.  In table 3, we can see that the variation in J is smaller 
 than those in H and K. The reason is that there are only  H and K data
 for the faint state (Glass 1981).  The variations in J and H  are the 
 same, and greater than in K if we do not consider the low state points. 

\subsubsection{B2 1308+326, $ OP 313, GC$}

 An emission line redshift of 0.996 comes from the identification of a fairly 
 strong, broad emission feature at 5586$\AA$ as MgII$\lambda$2800, while the 
 absorption redshift of 0.879 is from a weak doublet at $\lambda$5252.4$\AA$ 
 and 
 5264.4 $\AA$ (Miller et al. 1978). B2 1308+326 shows high and variable 
 polarization  indicating variability of $\Delta P = 10\%$ on a time scale 
 of 24 hours (Impey et al. 1984) and a decreasing of 12$\%$ in two days 
 (Moore et al. 1980).  The recorded maximum polarizations of 
 $P_{IR} = 19.6\%$ and $P_{opt} = (28\pm 5)\%$  at 0.44$\mu m$ was also 
 observed by Puschell et al. (1979). An optical variation 
 of $\Delta m$ = 5.6 mag has been observed (Angel \& Stockman, 1980). When the 
 infrared emission increases the spectral index does not change (Sitko et al.
 1983),  but an indication of a positive correlation between the infrared 
 spectral index  and brightness is found by Brown et al. (1989). The 
 1977/1978 optical outbursts  show a double peak structure separated by 
 1 year. Following the 1978 peak,  Puschell et al (1979) found that the 
 infrared fluxes did not decrease with time  as rapidly as the visual fluxes 
 and proposed that the more rapid decline  of visual compared with infrared 
 light is a general characteristic  of these events.  From the light curve,  
 we can see that the brightness decreases with time with some flux 
 fluctuations showing up. No definite relation can be obtained for the 
 spectral index and magnitude  for the long-term data, but there is a 
 correlation between (J-K) and (J-H): 
 (J-K) = 1.50(J-H) + 0.48 with $r$ = 0.596 and $p = 1.8 \times 10^{-3}$ 
 (see Fig. 15). 

\subsubsection{MC 1400+162, $4C 16.39, OO100$}

 This variable object ($\Delta m$ = 2.8, Zekl et al. 1981 ) is found in 
 a cluster  of galaxies (Miller 1978b) with a redshift of 
 $z_{em}$ $\sim$ 0.245  from the  [OII], $H_{\beta}$, [OIII], 
 [OII] (Miller et al. 1978) and show a  variable polarization 
 ($P_{opt}=4.4\%-14\%$, Angel \& Stockman 1980; Impey et al. 1984). The 
 infrared observations show a 0.54 mag variation  in K and 
  (H-K) = 0.89$\pm$0.11.  

\subsubsection{B2 1418+546, $OQ530$}

 1418+546 shows a large optical variation of 4.8 magnitude ranging from 
 11.3 to  16.1 magnitude in Miller's (1978) study; its  optical polarization 
 from 5$\%$ to 24$\%$ (Marchenko 1980) is greater than  the infrared 
 polarization of $P_{IR}=19\%$ (Impey et al. 1984). In the infrared, some 
 properties have been shown in the papers of O'Dell et al. (1978) and 
 Puschell \& Stein (1980). No correlation was found between  spectral 
 index and brightness, but the flux variations are found larger 
 at lower frequency by Massaro et al. (1995) and Takalo et al (1992). 
 For the spectral  shape and the brightness of the source,  the long-term 
 data show  (see Fig. 16):  (J-K) = -0.08H + 2.65 with 
  $r$ = -0.40 and $p = 2.6\%$, and (J-H) = -0.08K + 1.75 with $r$ = -0.41 
 and $p = 2.2\%$   indicating that the spectrum steepens when the source 
 brightens. No similar correlation has been found in the visual band 
 (Worrall et al 1984a).  The infrared light curves show two bursts separated 
 by about 8 years and the  variation at lower frequency to be greater.  
 The infrared variations are smaller than  in the optical band. 
 The (J-K) is found related with (J-H) and (H-K) (see Fig. 16):
 (J-K) = 1.56(J-H) + 0.43 with $r$ = 0.856 and $p = 1.6\times 10^{-8}$,
 (J-K) = 1.98(H-K) + 0.03 with $r$ = 0.61  and $p = 7.0\times 10^{-4}$.

\subsubsection{PKS 1514-241, $AP ~Librae$}

 This variable ( $\Delta m = 3.0$mag, Branly et al. 1996) object is located 
 at the center of an elliptical galaxy with an emission line redshift of 
 $z = 0.049$  (Disney et al. 1974), polarizations of $P_{3.7cm} = 4.4\%$ 
 (Wardle 1978) and  $P_{IR} = 2\%-7.4\%$ and $P_{opt} = 7\%$ (Impey et al. 
 1982) are reported.  The infrared data show  a correlation between (J-K) 
 and (H-K):  (J-K) = 0.18(H-K) + 0.84 with $r$ = 0.928 and 
 $p = 4.5\times 10^{-5}$.  For color  index and magnitude, there is an 
 indication of (J-K) increasing with H but this is far from any certain 
 conclusion (see Fig. 17).

\subsubsection{PKS 1519-273}

 This polarized (  $P_{opt} = 11.4\%$, Impey \& Tapia 1988) object has 
 only been
 observed  twice by Gear (1993) and Glass (1981).  The observations 
 indicate color indices of  (J-H) = 0.45$\pm$0.21, (J-K) = 1.35$\pm$0.17, 
 and (H-K) = 0.90$\pm$0.24.  The optical variation is $\Delta m$ = 2.43 
 (Fan 1997).

\subsubsection{PKS 1538+149,$4C14.60, OR165$}

 Polarizations of $P_{3.7cm} = 9.3\%$ (Wardle 1978) and $P_{opt} = 32.8\%$ 
 (Brindle et al 1986) are reported.  The infrared variations found here from  
 few data points  is much smaller than in the optical bands 
 ( $\Delta m$ = 3.7, Stickel et al. 1993). 

\subsubsection{B2 1652+398, $Mkn 501, 4C39.49, OS387$}

 B2 1652+398 was shown to have a redshift of 0.0337 by Ulrich et al. (1975). 
 It has effective spectral indices similar to that of  X-ray selected 
 BL Lac objects but has been classified as an RBLs. The largest optical 
 variation of  1.3 mag (Stickel et al. 1993) is smaller than in J. 
 Its optical and  infrared emissions are polarized at $P_{opt} = 2\%-7\%$ 
 and $P_{IR} = 3 \%$ ( CH(84)).  During the flare the colors are found to be
 very 
 much bluer than in quiescent state,  reaching (H-K) = 0.13 and (J-H)=0.48 
 (Takalo et al. 1992), but there are almost  no correlation between
  color index  and magnitude  or for color-color  indices (see Fig. 18).

\subsubsection{1727+502, $I ZW 187$}
 
 IZW 187 was observed spectroscopically by Oke (1978) to show a redshift of 
 0.0554. It is interesting that the optical polarization ($P_{opt} = 4\%-6\%$) 
 is smaller than the radio polarization ($P_{3.7cm} = 11.3\%$) (Kinman, 1976). 
 IZW 187 shows the highest ratio of X-ray luminosity to optical luminosity, 
 $L_{X-ray}/L_{opt.}$,  among BL Lac objects observed by $Einstein$ 
 observatory (Ku et al. 1982). An optical  variation of 2.1 mag was reported 
 by Scott (1976) (see also Bozyan et al. 1990).  Recently, it was 
 found to have a massive black hole ($5.4\times 10^{8}M_{\odot}$) at its 
 center  in our previous paper (Fan 1995).  Variations of $\Delta H $ = 
 0.57mag,  and $\Delta K $ = 0.82 mag are found. 
 There is an indication of color index (J-K) decreasing  with H. No  
 correlation for the color-color indices is really found (see Fig. 19). 

\subsubsection{PKS 1749+096, $OT 081, 4C 09.57$}

 This highly polarized ($P_{3.7cm} = 7.2\%$, Wardle 1978; $P_{opt} = 32 \%$, 
 Brindle et al. 1986 ) BL Lac object  was found to show an optical variation 
 of 2.7 mag  (Stickel et al. 1993). The infrared data show an indication of 
 a correlation  between J and (H-K) with  $r_{S}$ = 0.83 and 
 $p = 1.1\times 10^{-3}$. (J-K) is found correlated with (J-H) and (H-K): 
 (J-K) = 1.44(J-H) + 0.48 with 
 $r$ = 0.79 and $p = 2.4\times 10^{-3}$, (J-K) = 3.31 (H-K) -1.13 with 
 $r$ = 0.75 and $p = 5\times 10^{-3}$ (see Fig. 20).

\subsubsection{S5 1803+78}

 There are only observations of  $f_{J } = 5.1 mJy$, $f_{H} = 5.5  mJy$, 
 and $f_{K} = 6.7 mJy$,  which gives J = 13.68, H = 13.13, and K = 12.42 mag
 in the paper of Heckman et al. (1983) for this highly polarized 
 ($P_{opt} = 35.2\%$, Kuhr \& Schmidt, 1990) object.

\subsubsection{S4 1807+698, $3C371.0, 4C 69.24$}

 The N galaxy 3C371.0 ($z_{em}$ = 0.0506 Miller, 1975) is related to BL 
 Lac object  because of its optical variability ($\Delta m \sim $ 2.0 mag)
  and polarization  ($P \sim 12 \%$) (Shaffer 1978; Kinman 1976; CH(84)). 
 Miller (1975) suggests  that the polarization increases significantly in 
 the UV band if the visual polarization were diluted by 
 starlight. Capps \& Knacke (1978)  reported an expected lower-than-optical 
 infrared polarization ($P_{IR} = 5.3\pm1.5\%$)  when the starlight diluted 
 heavily the infrared.  There are only 7-night of infrared data in the 
 literature, giving  $f_{J} = 13.8  mJy$, 
 $f_{H} = 16.0 mJy$, and $f_{K} = 19.6 mJy$ (Heckman et al. 1983), included, 
 which show a variation of 1 mag in K (from 10.86 to 11.88 mag).

\subsubsection{S4 1823+568, $4C56.27$}

 There are only 2-night of infrared data for this polarized 
 ($P_{opt} = 16.8\%$, 
 Impey et al. 1991) object: J = 14.87-14.92, H = 13.92, and K = 13.02-13.08 
 (Gear 1993), averaged color indices of (J-H) = 0.97$\pm$0.03, 
 (J-K) = 1.85$\pm$0.07, and (H-K) = 0.87$\pm$0.04 can be obtained.  

\subsubsection{PKS 2005-489}

 This is a moderately variable object ($\Delta m$ = 0.53, Bozyan et al. 1990). 
 The limited infrared data only show a variation of about 0.25 mag. No 
 polarization is published.

\subsubsection{PKS 2131-021, $4C 02.81$}

 There are only two night of infrared data showing a variation of 0.8 mag 
 (J = 16.06-16.88, H = 15.05-15.97, K = 14.23-14.88),  which is comparable 
 with the optical variation of $\Delta m$ = 1.2 mag (Stickel et al. 1993). 
 Mean infrared color indices of (J-H) = 0.96$\pm$0.07, 
 (J-K) = 1.91$\pm$0.12,  and (H-K) = 0.96$\pm$0.16 can be obtained. 
 $P_{opt} = 16.9\%$ is reported by  Kuhr \& Schmidt (1990).

\subsubsection{PKS 2200+420, $VRO 42.22.01, BL Lacertae$}

 The prototypical object of its class lies in a giant elliptical galaxy at a 
 redshift of $\sim$ 0.07. Maximum radio polarization of $P_{3.7cm} = 6.4\%$ 
 (Wardle 1978), infrared polarization of $P_{IR} = 15.1 \%$ 
 (Impey et al. 1984)  and optical polarization of $P_{opt} = 23\%$ 
 (Kuhr \& Schmidt 1990) are obtained.  It has been observed for about 27 
 years in the infrared (see Fan et al. 1998c) and about 100 years in the 
 optical (Fan et al. 1998a). A 14-year period  and a maximum optical variation 
 of $\Delta B = 5.31$ are found from the B light curve (Fan et al. 1998a). 
 The optical variation  is greater  than infrared variation.  Correlations 
 have been found between (J-K) and (J-H)  and (H-K) but not between 
 color index and magnitude (Fig 21, also see Fan et al. 1998c).

\subsubsection{PKS 2240-260}

 There are only five-night of infrared data, showing a variation of 
 $\Delta K$ = 1.7 mag.  There is an indication of a correlation between 
 J-H and K, 
 but it should be  investigated with more observations. $P_{opt} = 15.1\%$ 
 is reported for  it (Impey \& Tapia 1988).

\subsubsection{PKS 2254+074, $OY 091$}

 The highly polarized object ($P_{3.7cm} = 8.1\%$, Wardle 1978, 
 $P_{opt} = 14\%-21\%,$  $P_{IR}=17.2\%$, Impey et al. 1982) shows an 
 optical variation of $\Delta m_{B}$ = 2.37 ( Bozyan et al. 1990). 
 The infrared data show correlations between color indices (see Fig.22g-i).

\subsubsection{2335+031,$4C03.59, OZ061$}

 There are only one night of infrared data for this object: J = 16.68, 
 H = 15.77, and  K = 14.89, which show color indices of (J-H) = 0.91$\pm$0.06,
 (J-K) = 1.79$\pm$0.06, and (H-K) = 0.88$\pm$0.06. An optical variation of 
 0.8 mag and a radio  polarization of $P_{R} = 6\%$ are reported (CH(84)).

\subsection{Summary}

 In this paper, the infrared variations are presented for 40 RBLs, the 
 light curves  are shown in Fig. 2 to 22 for 22 objects with more 
 observations. The amplitude variation in the optical and infrared 
 bands are compared. Correlations between color-color indices and between 
 color and magnitude are investigated: (J-K) is always found correlated
 with (J-H). For the color-magnitude relation, some objects ( 0048-097, 
 0735+178, 0851+202, 1215+303, 1219+285,  1749+096, and perhaps 0219+428, 
 0537-441, 1514-241) show a color index increasing with magnitude, indicating
 that the spectrum flattens when the source brightens while some other objects
 ( 0754+100, 1147+245, 1418+546, and 1727+502) perhaps show an opposite 
 behaviour, the rest objects do not show any clear tendency between color 
 index and magnitude.

\begin{acknowledgements}{The authors are grateful to the anonymous referee 
 for his/her comments and to Dr. G. Adam for his critical reading. J.H. F
 thanks Y. Copin for his help with the figures. This work 
 is supported by the National Pan Deng  Project of China} 
\end{acknowledgements}
\cite{}

\begin{small}
\begin{table}
\caption[]{Literature and telescopes for the data}
\begin{tabular}{cc}
\hline\noalign{\smallskip}
Observer(s) &  Telescope(s)\\ 
\hline\noalign{\smallskip}
Allen(1976)             &  UM/UCSD 1.5m \\  
Allen et al (1982)      &  A-A 3.9m; UKIRT 3.8m\\  
Andrews et al(1974)     & SAAO 1.02m \& 0.52m \\  
Bersanelli et al (1992) &  ESO 3.6m \& 2.2m\\ \ 
Bregman et al (1984)    & Mount Lemmon 1.5m \\ \ 
Bregman et al (1990)    & UKIRT 3.8m; Hale 5m \\  
Brindle et al (1986)    &  UKIRT 3.8m \\  
Brown et al (1989, 1990)      &  UKIRT 3.8m \\  
Capps \& Knacke (1978)  &  KPNO 2.1m \& 4m\\  
Cruz-Gonzales \& Huchra (1984) & CTIO 4m; KPNO 1.5m \\  
Epstein et al (1972)    & 100-inch \\  
Falomo et al (1988)     &  Carlos Sanchez (TCS) 1.5m \\  
Falomo et al (1993)     &  ESO 2.2m \\  
Gear      (1993)       &  UKIRT 3.8m  \\  
Gear et al (1985,1986a,b)       &  UKIRT 3.8m  \\  
Glass (1981)            &  Sutherland 1.88m \\  
Heckman et al(1983)     & UAO 2.25m, 1.54m \\  
Holmes et al (1984a,b)     &  UKIRT 3.8m  \\  
Impey et al (1982,1984)      &  UKIRT 3.8m \\  
Impey (1983)            &  Hawaii 2.2m \\  
Kawai et al(1991)       & UKIT,UMIT \\  
Kidger et al (1994)     & TCS 1.5m            \\  
Kotilainen et al (1992) & UKIRT 4m \\  
Knacke et al (1976)     & PKBT 2.1m \\ \hline 

\end{tabular}
\end{table}

\begin{table}
\caption[]{Literature and telescopes for the data}
\begin{tabular}{cc}
\hline\noalign{\smallskip}
Observer(s) &  Telescope(s)\\ 
\hline\noalign{\smallskip}
Landau et al (1986)     &  UKIRT 3.8m; Hale 5m \&\\
                        &   Mount Lemnon 1.5m \\  
Lepine et al(1985)      &  ESO 3.6m \\  
Litchfield et al (1994) &  ESO 2.2m  \\  
Lorenzetti et al(1989)  &  Italian 1.5m\\  
Massaro et al (1995)    & TIGRO 1.5m \& ESO 1.0m \\  
Mead et al (1990)       &  UKIRT 3.8m\\  
Moore et al (1980,1982)      &  2.29m; 1.54m;UAO 1.5m\\  
O'Dell et al (1977,1978a,b)     &  UM/UCSD 1.5m \\  
Puschell et al (1979, 1983)   &  UM/UCSD 1.5m \\  
Puschell \& Stein (1980)&  UM/UCSD 1.5m \\  
Rieke et al (1976)      & KPNO 4m NASA 1.54m \\       
                        & Steward Obs. 2.25m        \\  
Rieke et al (1977)      &  UOA 90inch \& 61 inch \\  
Roelling et al (1986)    &  UKIRT 3.8m \\  
Sitko et al (1983)     & UM/UCSD 1.5m  \\  
Sitko \& Sitko (1991)   &  KPNO 1.3m \& 1.5m \\  
Smith et al (1987)      &  KPNO 2.1m \\  
Takalo et al(1992)      & TCS 1.5m\\  
Tanzi (1986)            &  ESO 3.6m \\  
Tanzi et al (1989)      &  ESO 1.5m \& 3.6m \\  
Ulrich et al (1975)      & Steward 2.25m \& NASA 1.52m \\  
Wolstencroft et al (1982) & UKIRT 3.8m \\  
Worrall et al (1984a,1986)    &  MU/UCSD 1.5m \\  
Worrall et al (1984b)    &  Shane 3m; MU/UCSD 1.5m \\  
Worrall et al (1984c)    & Lick Obs. 3m; Hale 5m \\  
Xie et al (1994)        & Xing Long, BAO 1.26m \\  \hline
\end{tabular}
\end{table}

\begin{table*}
\caption[]{Observed Largest Variations of BL Lac Objects}
\begin{tabular}{cccccccccc}
\noalign{\smallskip}
\hline
\noalign{\smallskip}
 Name     & z & $A_{V}$ & $P_{opt}$ & $\Delta m_{opt}$ & Infrared
Variation   & Averaged
Color Index \\      \hline
          &           &      & ($\%$) &      & $\Delta$ J &  $\Delta$ H & $ \Delta$ K   &  J-H    & H-K    \\  \hline
0048-097  & ------    & 0.0  & 27.  & 2.7  & 6.55       &  6.23         & 6.01        & 0.88 $\pm$ 0.18  & 0.85 $\pm$ 0.10 \\  
0118-272  & 0.557     & 0.0  & 17.0 & ---- & 0.56       &  0.57         & 0.67        & 0.83 $\pm$ 0.04  & 0.82 $\pm$ 0.02 \\  
0138-097  & 0.50      & 0.0  & 29.3 & 1.52 & 1.69       &  1.57          & 1.56       &0.84  $\pm$ 0.08  & 0.80 $\pm$ 0.03 \\   
0219+428  & 0.444     & 0.51 & 18.0 & 2.0  & 1.57       &  1.58          & 1.61       & 0.76 $\pm$ 0.14  & 0.82 $\pm$ 0.12 \\   
0235+164  & 0.940     & 0.10 & 44.0 & 5.3  & 5.0        & 4.87           & 4.93       & 0.96 $\pm$ 0.18  & 1.02 $\pm$ 0.17 \\   
0537-441  & 0.896     & 0.19 & 18.7 & 5.4  & 3.00       & 2.75           & 2.43      & 0.87  $\pm$ 0.08  & 0.95 $\pm$ 0.09\\   
0735+178  & 0.424     & 0.46 & 35.0 & 4.6  & 2.47       &  2.30          & 2.40       & 0.86 $\pm$ 0.11  & 0.86 $\pm$ 0.12 \\  
0754+100  & 0.66      & 0.40 & 26.0 & 2.0  & 1.77       & 2.05           & 1.88       & 0.82 $\pm$ 0.12  & 0.84 $\pm$ 0.10 \\  
0818-128  & ------    & 0.64 & 36.0 & 2.9  & 2.14       & 2.14           & 2.00       & 0.82 $\pm$ 0.12  & 0.81 $\pm$ 0.08 \\  
0829+046  & 0.18      & 0.31 & 12.0 & 5.0  & 2.11       & 2.08           & 2.15       & 0.91 $\pm$ 0.17  & 0.84 $\pm$ 0.11 \\  
0851+202  & 0.306     & 0.13 & 37.2 & 6.0  & 3.87       & 3.78           & 3.54       & 0.82 $\pm$ 0.16  & 0.88 $\pm$ 0.14 \\  
1101+384  & 0.03      & 0.0  & 7.9  & 4.6  & 1.03       & 0.98           & 2.03       & 0.64 $\pm$ 0.07  & 0.54 $\pm$ 0.06\\   
1144-379  & 1.048     & 0.32 & 8.5  & 1.92 & 3.17       & 3.43           & 3.46       & 0.87 $\pm$ 0.09  & 0.97 $\pm$ 0.07 \\  
1147+245  & ------    & 0.0  & 13.0 & 0.77 & 1.16       & 1.34           & 1.12       & 0.81 $\pm$ 0.15  & 0.85 $\pm$ 0.11 \\  
1215+303  & 0.237     & 0.0  & 14.0 & 3.3  & 1.05       & 0.78           & 0.89       & 0.71 $\pm$ 0.11  & 0.63 $\pm$ 0.09 \\  
1219+285  & 0.13      & 0.0  & 10.0 & 5.0  & 1.01       & 2.46           & 2.46       & 0.79 $\pm$ 0.14  & 0.76 $\pm$ 0.11 \\  
1308+326 & 0.996      & 0.0  & 28.0 & 5.6  &  5.07      & 4.80           & 5.55       & 0.84 $\pm$ 0.21  & 0.90 $\pm$ 0.19\\   
1418+546  & 0.152     & 0.0  & 24.0 & 4.8  &  1.11      & 1.22           & 1.59       & 0.82 $\pm$ 0.09  & 0.81 $\pm$ 0.09 \\  
1514-241  & 0.049     & 0.24 & 7.0  & 3.0  &  1.40      & 1.45           & 2.64      & 0.82  $\pm$ 0.09  & 0.68 $\pm$ 0.09\\   
1538+149  & 0.605     & 0.01 & 32.8 & 3.7  &  0.53      &  0.89          & 0.65      & 0.84  $\pm$ 0.08  & 0.84 $\pm$ 0.10 \\  
1652+398  & 0.033     & 0.10 & 7.0  & 1.30 & 2.37       & 1.67           & 1.40       & 0.74 $\pm$ 0.13  & 0.39 $\pm$ 0.12\\   
1727+502  & 0.055     & 0.16 & 6.0  & 2.1  & 0.55       &  0.70          &  0.82      & 0.63 $\pm$ 0.05  & 0.44 $\pm$ 0.03\\   
1749+096  & 0.320     & 0.48 & 32.0 & 2.7  & 2.16       & 2.21           & 1.96       & 0.96 $\pm$ 0.11  & 0.89 $\pm$ 0.08 \\  
1807+698  & 0.051     & 0.21 & 12.0 & 2.0   &  0.48      &  0.79          &  1.02      & 0.61 $\pm$ 0.16  & 0.56 $\pm$ 0.09\\   
2005-489  & 0.071     & 0.17 & ---  & 0.53 &  0.31      &  0.29          &  0.26      & 0.69 $\pm$ 0.02  & 0.59 $\pm$ 0.02\\   
2200+420  & 0.069     & 0.97 & 23.0 & 5.31 & 2.29       & 2.42           &   2.93     & 0.89 $\pm$ 0.11  & 0.85 $\pm$ 0.12 \\  
2240-260  & 0.774     & 0.0 &  15.1 & ---  & 1.28       & 1.01           &  1.66      & 0.88 $\pm$ 0.10  & 0.94 $\pm$ 0.07 \\  
2254+074  & 0.190     & 0.04 & 21.0 & 2.37 & 0.97       & 1.36           &  1.0       & 0.83 $\pm$ 0.09  & 0.75 $\pm$ 0.05 \\  \hline 
\end{tabular}
\end{table*}
\end{small}

\newpage
\begin{table*}
\caption[]{Near-Infrared Observations of 40 Radio-Selected BL Lac Objects}
\begin{tabular}{cccccccc}
\hline
\noalign{\smallskip}
 Name &  JD 2400000+ & J  & $\Delta$ J  & H  & $\Delta$ H & K  & $\Delta$ K \\
 (1)  &  (2)         & (3)& (4)         &(5) & (6)        &(7) & (8)  \\ 
\hline
\end{tabular}

\end{table*}

\newpage 
Figure Captions
	 
Fig.  1.  Plot of averaged color indices, (J-H) versus (H-K), the error bars
 are $1\sigma$ deviations\\
Fig.  2. Light curves and color index properties for 0048-097. $a$: J light curve; $b$: H light curve; $c$: K light curve;
 $d$: J-H vs. K; $e$: J-K vs. K; $f$: H-K vs K; $g$: H-K vs. J-H; $h$: J-K vs. J-H; $i$: J-K vs. H-K\\
Fig.  3. Light curves and color index properties for 0219+428\\
Fig.  4. Light curves and color index properties for 0235+164\\
Fig.  5. Light curves and color index properties for 0537-441\\
Fig.  6. Light curves and color index properties for 0735+178\\
Fig.  7. Light curves and color index properties for 0754+100\\
Fig.  8. Light curves and color index properties for 0818-128\\
Fig.  9. Light curves and color index properties for 0829+046\\
Fig. 10. Light curves and color index properties for 0851+202\\
Fig. 11. Light curves and color index properties for 1101+384\\
Fig. 12. Light curves and color index properties for 1147+245\\
Fig. 13. Light curves and color index properties for 1215+303\\
Fig. 14. Light curves and color index properties for 1219+285\\
Fig. 15. Light curves and color index properties for 1308+326\\
Fig. 16. Light curves and color index properties for 1418+546\\
Fig. 17. Light curves and color index properties for 1514-24 \\
Fig. 18. Light curves and color index properties for 1652+398\\
Fig. 19. Light curves and color index properties for 1727+502\\
Fig. 20. Light curves and color index properties for 1749+096\\
Fig. 21. Light curves and color index properties for 2200+420\\
Fig. 22. Light curves and color index properties for 2254+074
	 
\end{document}